\documentclass[aps,10pt,prx,twocolumn,showpacs,floatfix,superscriptaddress,longbibliography,nofootinbib
]{revtex4-2}

\usepackage{quantikz}
\usepackage{times, ulem}
\usepackage{graphicx}
\usepackage{graphics}
\usepackage{dcolumn}
\usepackage{bm}
\usepackage{amsmath, amssymb}
\usepackage{amssymb}
\usepackage{hyperref}
\hypersetup{colorlinks=true,linkcolor=blue,citecolor=blue,urlcolor=blue}
\usepackage[export]{adjustbox}

\begin{document}

\title{Information recycling in coherent state discrimination}
\author{L. F. Melo}
\affiliation{QuIIN - Quantum Industrial Innovation, Centro de Competência EMBRAPII CIMATEC, SENAI CIMATEC, Av. Orlando Gomes, 1845, Salvador, BA 41850-010, Brazil}
\email{lucas.felipe@fbter.org.br}
\author{János A. Bergou}
\affiliation{Department of Physics and Astronomy, Hunter College of the City University of New York, 695 Park Avenue, New York, NY 10021, USA}
\author{Alexandre B. Tacla}
\affiliation{QuIIN - Quantum Industrial Innovation, Centro de Competência EMBRAPII CIMATEC, SENAI CIMATEC, Av. Orlando Gomes, 1845, Salvador, BA 41850-010, Brazil}
\email{alexandre.tacla@fieb.org.br}
 
\date{\today}

\begin{abstract}
The discrimination of coherent states is a crucial component in quantum communication with continuous variables, especially in quantum key distribution protocols (CV-QKD), which rely on the ability to distinguish among different coherent states to establish a shared secret key between two parties. Here, we propose and analyze a strategy for distinguishing among $N$ phase-symmetric coherent states, which optimally takes unambiguous discrimination (UD) to the deterministic regime, at the inevitable cost of having non-zero probability of error. Despite the disturbance introduced by the separation map used in the UD process, we show that for $N>2$, the ``failure'' states of UD retain residual information about the original input states, which can be further used for discrimination. Rather than discarding inconclusive outcomes as in conventional UD, we show that the ``failure'' states of UD can be optimally recycled by performing a sequential minimum-error discrimination (MED). This strategy, which we call \textit{information recycling} (IR), combines the benefits of both MED and optimal UD: It always provides conclusive results while allowing for a subset of those results to be error-free, which are identifiable by an ancillary system. We characterize the disturbance introduced by the state separation map by the infidelity between input and failure states, demonstrating that it lower bounds the error probability in the recycling stage. Furthermore, in the low-amplitude regime--relevant for long-distance CV-QKD applications--we show that the state separation achieves significant success while introducing relatively low disturbance to the input states after failed events. Our results open up new possibilities for adaptive and sequential discrimination protocols in continuous-variable settings, and could potentially be used in the design of next-generation receivers in quantum communication.
\end{abstract}

\maketitle

\section{Introduction}

Coherent states of the quantum electromagnetic field are important carriers of information in quantum communication protocols and have gained special interest in recent decades due to their compatibility with commercially available optical telecommunication components~\cite{Huttner95, Banaszek99, Leverrier11, Lin19, Zhang24}. In such systems, a set of coherent states is often used to encode and transmit information, where the efficiency of information retrieval directly depends on the receiver's ability to discriminate them. Consequently, quantum measurements capable of efficiently distinguishing between coherent states have been crucial ingredients in applications of continuous-variable quantum cryptography~\cite{Huttner95, Banaszek99, Dusek00, Enk02, Izumi21}, quantum sensing~\cite{Giovannetti04, Ranjith12, Primaatmaja21}, and quantum key distribution~\cite{Wittmann10, Liao18, Notarnicola23-2}. However, coherent state discrimination becomes particularly challenging when dealing with states that differ by only a few photons on average. In such cases, the states are highly non-orthogonal and this task cannot be accomplished both deterministically and without error~\cite{Helstrom69, Chefles98-2}.

The main optimal strategies to distinguish non-orthogonal quantum states are the minimum-error discrimination (MED), for which the average probability of error is minimized given that a conclusive result is always obtained~\cite{Helstrom69, Ban97}, and the optimal unambiguous discrimination (UD), for which error-free results are obtained at the expense of having inconclusive results at the lowest possible rate~\cite{Ivanovic87, Dieks88, Peres88, Chefles98-1}. This can be accomplished by a probabilistic separation map that takes the non-orthogonal input states to orthogonal ones with optimal success probability~\cite{Prosser16, Chefles98, Nakahira12}; when this separation map fails, the inconclusive output states are discarded. However, when the failure probability is large, the optimal UD may become too inefficient for practical purposes. This is typically the case, for instance, in the discrimination of phase-symmetric coherent states, which are commonly used in continuous-variable quantum key distribution (CV-QKD) protocols~\cite{Leverrier11, Lin19, Liao18, Notarnicola23-2}. In such a case, the performance of optimal UD decreases with an increasing number of states and decreasing amplitude~\cite{Chefles98-2}, making optimal UD unfavorable as a receiver strategy for long-distance CV-QKD with large alphabets.

In this paper, we revisit the problem of optimal UD of $N$ equiprobable phase-symmetric coherent states equally sampled. We show that for $N>2$, the ``failure'' states of optimal UD still retain residual information about the input states, which can be further used for discrimination. While these failure states cannot provide further unambiguous identifications, they retain residual information that can be optimally extracted through sequential MED. Rather than discarding inconclusive outcomes, this recycling strategy systematically recovers information that would otherwise be lost. This strategy, which we henceforth call \textit{information recycling} (IR), increases the average probability of correct identifications by optimally taking UD to the deterministic regime, while allowing for a subset of those results to be error-free, which are identifiable by an ancillary system. Building on Refs.~\cite{Roa11, Zhang14, Melo25}, which introduced information recycling for specific discrete-variable systems (equidistant states, qutrits, and symmetric qudit states, respectively), we extend this idea to continuous variables by developing a general IR framework for arbitrary-size sets of phase-symmetric coherent states. 

We develop a comprehensive analytical framework for IR by decomposing optimal UD into a state separation map followed by projective measurement. This decomposition enables us to derive the failure states analytically and characterize the reduced dimensionality of their subspace. We quantify IR performance through both the optimal unambiguous success probability and the correct identification probability for failure state discrimination, demonstrating that the state separation introduces modest disturbance in the low-amplitude regime. Lastly, we analyze the information gain of IR over optimal UD, by calculating the classical mutual information for both strategies. We show that recycling yields substantial mutual information gain over standard UD that increase systematically for larger alphabets. These insights could potentially be used in the design of receivers for quantum communication tasks where efficient and practical state discrimination is critical.

This paper is organized as follows. We start by providing, in Sec.~\ref{sec:theory}, a brief overview of the main optimal state discrimination strategies: MED and optimal UD. In Sec.~\ref{sec:inforec}, we use these results to build the IR strategy. We start by determining its POVM operators and the detection probabilities. Then, we find the failure states of the state separation map. Finally, we quantify the strategy performance through complementary probabilistic and information-theoretic analyses. We present our conclusions and perspectives for future research in Sec.~\ref{sec:conclusion}.


\section{Optimal coherent state discrimination}\label{sec:theory}

In this section, we present a brief overview of the main optimal strategies for discriminating $N$ phase-symmetric coherent states: Minimum-error discrimination (MED) and optimal unambiguous discrimination (UD). We assume that all states can be prepared with equal \textit{a priori} probability. This family of states is of particular interest in quantum communication~\cite{Leverrier11, Ranjith12, Liao18, Primaatmaja21, Sidhu23, Notarnicola23-2} and admits analytical solutions for both discrimination strategies. 

The problem consists in optimally deciding which hypothesis best describes a quantum system that may be in one of the states from a known set $\lbrace |\alpha_k\rangle \rbrace_{k=0}^{N-1}$. This requires performing a quantum measurement and using the outcome $k'$ to infer that the state of the system was $|\alpha_{k'}\rangle$. This measurement is described by a positive operator-valued measure (POVM) (a set of positive semidefinite operators $\lbrace \hat{\Pi}_j\rbrace$ that satisfy $\sum_j \hat{\Pi}_j = \hat{I}$
) and its optimality depends on the figure of merit of interest: The average probability of correct results is maximized by MED, whereas the individual confidence in the results is maximized by optimal UD. We note that both strategies require non-Gaussian measurements, which outperform homodyne and heterodyne schemes~\cite{Giovannetti04, Takeoka08, Cariolaro15, Roberson21}. Practical implementations include receivers based on linear optics and photodetection~\cite{Takeoka06, Wittmann10, Sych16, DiMario18, Thekkadath21, Notarnicola23, Enk02, Sedlak07, Muller12, Izumi12, Becerra13, Becerra13-2, Izumi21, Sidhu23}, probabilistic amplification~\cite{Pandey13, Rosati16}, and ancilla-based measurements~\cite{Han18-2, Han20}.

\subsection{Phase-symmetric coherent states} \label{subsec:symmetric}

Deriving general analytical solutions for the main optimal discrimination strategies for arbitrary sets of coherent states is a difficult task. The general approach to overcome this problem involves using semidefinite programming techniques (see Ref.~\cite{Primaatmaja21}). Among analytically solvable cases, phase-symmetric sets of coherent states are particularly important for continuous-variable quantum communication~\cite{Primaatmaja21, Leverrier11, Lin19} and sensing~\cite{Ranjith12}, as they enable feasible practical implementations \cite{Leverrier11, Lin19, Notarnicola23-2}. This family of states, defined in the following, admits analytical expressions for the POVM operators of the main optimal discrimination strategies for sets of any size, as long as its elements are equiprobable~\cite{Chefles98-2, Ban97}.

Let $\lbrace |\alpha_k\rangle\rbrace_{k=0}^{N-1}$ be a set of $N$ phase-symmetric coherent states, which can be written in the Fock basis $\lbrace |n\rangle\rbrace$ as
\begin{align}    \label{eq:coherent}
    |\alpha_k\rangle = e^{-\alpha^2/2} \sum_{n=0}^{\infty} \frac{\alpha^n \omega^{kn}}{\sqrt{n!}} |n \rangle,
\end{align}
where $\omega\equiv\exp\left(2\pi i/N\right)$. Coherent states are not mutually orthogonal and, in the phase-symmetric case, have their inner product given by
\begin{align}   \label{eq:innerprod}
    \langle \alpha_j|\alpha_k\rangle = e^{\alpha^2(\omega^{(k-j)}-1)},
\end{align}
which only vanishes in the limit $\alpha\rightarrow \infty$. Without loss of generality, we henceforth assume that $\alpha$ is a real number and write the states~(\ref{eq:coherent}) in an orthonormal basis as
\begin{align}    \label{eq:alphacj}
    |\alpha_k\rangle = \sum_{j=0}^{N-1} c_j \omega^{kj} |\phi_j\rangle,
\end{align}
where the states $\lbrace|\phi_j\rangle \rbrace_{j=0}^{N-1}$ are defined by (see Appendix~\ref{app:orthonormal} for details)
\begin{align}    \label{eq:phij}
    |\phi_j\rangle = \frac{e^{-\alpha^2/2}}{c_j} \sum_{p=0}^{\infty} \frac{\alpha^{j+pN}}{\sqrt{(j+pN)!}} |j+pN\rangle,
\end{align}
and the coefficients $\lbrace c_j \rbrace_{j=0}^{N-1}$ are given by~\cite{Chefles98-2}
\begin{align}   \label{eq:coefficients}
   c_j^2 = \frac{1}{N} \sum_{\ell=0}^{N-1} \omega^{-j \ell} e^{\alpha^2(\omega^\ell-1)}.
\end{align}
Since the states $\lbrace |\phi_j\rangle\rbrace$ are orthonormal, the states~(\ref{eq:alphacj}) span an $N$-dimensional subspace and thus are linearly independent~\cite{Chefles98-2}. It is worth pointing out that the coefficients $c_j$ are the same for all $|\alpha_k\rangle$, which only differ by their relative phase. In Appendix~\ref{app:orthonormal}, we show plots of Eq.~(\ref{eq:coefficients}) in terms of the mean photon number for sets with $N=3,4,5,6$ coherent states.

\subsection{Minimum-error discrimination}

An essential figure of merit for general discrimination problems is the average probability of having correct identifications. For $N$ equiprobable states, this probability is given by
\begin{align}   \label{eq:probcorrect}
    P_c = \frac{1}{N} \sum_{k=0}^{N-1} \textrm{Tr} (\hat\Pi_k\hat{\rho}_k),
\end{align}
where $\hat{\rho}_k = |\alpha_k\rangle \langle\alpha_k|$ corresponds to the density matrix of the $k$th state of the input set and $\{\hat\Pi_k\}$ are POVM operators. The maximization of Eq.~(\ref{eq:probcorrect}) leads to the so-called minimum-error discrimination (MED). For $N$ phase-symmetric equiprobable states, the MED is represented by a POVM with operators~\cite{Ban97, Kato99, NotarnicolaThesis}
\begin{align}   \label{eq:ME_POVM}
     \hat{\Pi}^{\textsc{med}}_k = |u_k\rangle\langle u_k|,
\end{align}
where
\begin{align} \label{eq:uk}
    |u_k\rangle \equiv \frac{1}{\sqrt{N}} \sum_{j=0}^{N-1}\omega^{kj} |\phi_j\rangle.
\end{align}
The POVM elements~(\ref{eq:ME_POVM}) correspond to orthogonal projective measurements and satisfy the completeness relation $\sum_{j=0}^{N-1} \hat{\Pi}^{\textsc{med}}_j = \sum_{j=0}^{N-1} |\phi_j\rangle\langle\phi_j| = \hat{I}$, where $\hat{I}$ is the identity operator on the $N$-dimensional Hilbert space spanned by $\{|\phi_j\rangle\}$. Then the average probability of having a correct identification is determined by the coefficients~(\ref{eq:coefficients})
\begin{align}   \label{eq:Helstrom}
    P_c^\textsc{med} = \frac{1}{N}\left(\sum_{j=0}^{N-1} c_j \right)^2.
\end{align}
This quantity represents the fundamental limit imposed by quantum mechanics on state discrimination--known as the Helstrom bound~\cite{Helstrom69,Ban97}--here applied to the case of phase-symmetric equiprobable states.

\subsection{Optimal unambiguous discrimination}

Despite maximizing the average probability of correct identifications, all individual results obtained in the MED are ambiguous in the sense that one cannot infer with certainty which was the signal state based on the measurement outcome. An alternative strategy is to consider unambiguous discrimination (UD), which allows one to obtain error-free results, but only probabilistically ~\cite{Ivanovic87, Dieks88,Peres88,Chefles98-1}.

Given that the ensemble state is 
\begin{align}
    \hat{\rho} = \frac{1}{N}\sum_{k=0}^{N-1} |\alpha_k\rangle \langle\alpha_k|
\end{align}
and that one obtains a measurement result $k'$, the probability that the signal state was $|\alpha_k\rangle$ is determined by Bayes' rule
\begin{align}  \label{eq:Bayes}
    p(\alpha_k|k') = \frac{p(k'|\alpha_k)}{N p(k')},
\end{align}
where $p(k'|\alpha_k)=\textrm{Tr}(\hat{\Pi}_{k'}\hat{\rho}_k)$ is the probability of obtaining the outcome $k'$ given that the signal state was $|\alpha_k\rangle$, and $p(k')=\textrm{Tr}(\hat{\Pi}_{k'}\hat{\rho})=1/N\sum_{k=0}^{N-1}p(k'|\alpha_k)$ is the prior probability of obtaining the outcome $k'$. UD maximizes individual confidence, which is defined as the probability of identifying that the prepared state was $|\alpha_k\rangle$ given that the measurement result was $k'=k$
\begin{align}   \label{eq:confidence}
    C_k \equiv p(\alpha_k|k).
\end{align}
For UD, $C_k=1$ for all $k$. As the states to be discriminated are non-orthogonal, the operators $\lbrace \hat{\Pi}_j \rbrace_{j=0}^{N-1}$ cannot form a POVM alone while fulfilling the condition of providing error-free results $\textrm{Tr}(\hat{\Pi}_j \hat{\rho}_k) = 0$ for all $k\neq j$. Thus, an additional inconclusive result operator $\hat{\Pi}_?=\hat{I}-\sum_{j=0}^{N-1} \hat{\Pi}_j$ is needed to form a complete POVM set.

Optimal UD is obtained by minimizing the probability of obtaining inconclusive results $P_? = \textrm{Tr}(\hat{\Pi}_? \hat{\rho})$. Analytical solutions to this problem can be found in Refs.~\cite{Ivanovic87, Dieks88, Peres88, Chefles98-1, Chefles98-2, Bergou12}. In particular, for the case of phase-symmetric coherent states, the optimal success probability is known to be given by~\cite{Chefles98-2}
\begin{align}   \label{eq:Ps}
    P_s = N c_\textrm{min}^2,
\end{align}
where $c_\textrm{min}=\textrm{min}\lbrace c_j|j=0,\dots,N-1 ~\wedge~ c_j\neq 0 \rbrace$ [see Eq.~(\ref{eq:coefficients})]. As the expression for the coefficients $c_j$ cannot be further simplified for arbitrary $N$, we search for the minimum coefficient numerically, in order to calculate the probability (\ref{eq:Ps}). An alternative approach is to consider the method proposed in~\cite{Bergou12} to find analytical solutions to $P_s$. In Appendix~\ref{app:analytical}, we discuss a continuous-variable version of this method and find the solution for the three-state problem as an example.

\subsubsection{Optimal state separation}

As shown in Refs.~\cite{Chefles98, Nakahira12}, optimal UD can be seen as a two-step process. The first step consists of an optimal state separation map, which transforms the non-orthogonal input states~(\ref{eq:coherent}) into orthogonal distinguishable states~(\ref{eq:uk}) with optimal success probability.\footnote{Such a transformation can only be performed probabilistically; otherwise, one would be able to perfectly discriminate between non-orthogonal states.} If the separation operation is successful, a follow-up projective measurement discriminates the output states without error. Otherwise, the failure states are discarded. Optimal state separation was analytically solved for symmetric sets of pure equiprobable discrete-variable states of arbitrary dimension in~\cite{Prosser16}. Using the decomposition in Eq.~(\ref{eq:alphacj}), we are able to derive an analogous approach to find a solution for phase-symmetric coherent states.

The optimal state separation map can be implemented by a unitary operator that acts on the quantum system and on a two-dimensional ancilla, whose Hilbert space $\mathcal{H}_\mathrm{a}$ is spanned by the basis $\lbrace |0\rangle, |1\rangle \rbrace$, 
\begin{align}
    \hat{\mathcal{U}} |\alpha_k\rangle|1\rangle
    &= \hat{A}^s|\alpha_k\rangle |1\rangle+\hat{A}^f |\alpha_k \rangle|0\rangle \nonumber \\ 
    &= \sqrt{P_s}|\psi_k\rangle |1\rangle+\sqrt{1-P_s} |\beta_k \rangle|0\rangle, \label{eq:unitary}
\end{align}
where $\hat{A}^s,\hat{A}^f$ are the Kraus operators associated with success and failure events, respectively, $\lbrace|\psi_k\rangle \rbrace_{k=0}^{N-1}$ is an arbitrary set of orthonormal states, and $\lbrace |\beta_k\rangle \rbrace_{k=0}^{N-1}$ corresponds to the set of failure states. The choice of the initial state of the ancillary system is arbitrary; here we initialize the ancilla to $|1\rangle$ without loss of generality~\cite{Prosser16}. The state separation map is then concluded by performing a projective measurement on the ancilla; here, the result $1$ ($0$) indicates a successful (failed) separation. This scheme transforms the non-orthogonal input states~(\ref{eq:coherent}) into orthogonal states with optimal success probability $P_s$ given by Eq.~(\ref{eq:Ps}). For convenience, which will become clear in Sec.~\ref{sec:inforec}, we choose $|\psi_k\rangle=|u_k\rangle$, defined in Eq.~(\ref{eq:uk}). In this way, the success states will have the same symmetry of the input states~(\ref{eq:alphacj}), and the final measurement to discriminate the states with optimal UD will coincide with the minimum-error projectors defined in Eq.~(\ref{eq:ME_POVM}). Under this assumption, state separation is implemented by the following Kraus operators:
\begin{subequations}  \label{eq:KrausSep}
\begin{align}
    \hat{A}^s &= \sum_{j=0}^{N-1} \frac{c_\textrm{min}}{c_j}  |\phi_j\rangle\langle \phi_j| \label{eq:SepAs}\\
    \hat{A}^f &= \hat{U}_f \sum_{j=0}^{N-1} \sqrt{1 - \left(\frac{c_\textrm{min}}{c_j}\right)^2} |\phi_j\rangle\langle \phi_j|, \label{eq:SepAf}
\end{align}
\end{subequations}
where the unitary $\hat{U}_f$ comes from the polar decomposition of the operator associated with failure events, $\hat{A}^f=\hat{U}_f \sqrt{\hat{\Pi}^f}$, with $\hat{\Pi}^f=\hat{I}-\hat{\Pi}^s$ and $\hat{\Pi}^s= \hat{A}^{s~\dagger} \hat{A}^s$. We are free to choose the unitary $\hat{U}_f$, which reflects the freedom to choose the physical setup, and the form of the failure sets $\lbrace |\beta_k\rangle \rbrace_{k=0}^{N-1}$ depends on this choice.

Optimal UD is then represented by the ($N+1$)-outcome POVM determined by elements $\lbrace \hat{\Pi}^s_0, \ldots, \hat{\Pi}^s_{N-1}, \hat{\Pi}^f \rbrace$, where $\hat{\Pi}^s_j = \hat{A}_j^{s~\dagger} \hat{A}^s_j$ with Kraus operator defined by
\begin{align}     \label{eq:IRAs}
    \hat{A}^s_j = \sqrt{\hat{\Pi}_j^{\textsc{med}}} \hat{A}^s.
\end{align}
It can be easily verified that $\hat{\Pi}^f  + \sum_{j=0}^{N-1} \hat{\Pi}^s_j = \hat{I}$. In the next section, we show how this measurement can be slightly changed to extract potentially useful information from the states $\lbrace |\beta_k\rangle \rbrace_{k=0}^{N-1}$.

\section{Discrimination with information recycling}\label{sec:inforec}

The failure states $\lbrace |\beta_k\rangle \rbrace_{k=0}^{N-1}$ of the optimal UD are conventionally discarded as inconclusive results since they cannot be used to extract more unambiguous identifications of the input states $\lbrace |\alpha_k\rangle \rbrace_{k=0}^{N-1}$. However, previous papers have pointed out that, for discrete-variable systems, correct identifications of the input states could still be extracted from the failure set with reduced confidence. Pure equidistant states were considered in~\cite{Roa11} and a few examples of pure qutrit states in~\cite{Zhang14}. Recently, a MED of the failure states of optimal UD was experimentally demonstrated for three pure symmetric qutrit states~\cite{Prosser22}; an analytical solution for pure symmetric qudit states was achieved by one of the present authors~\cite{Melo25}.

Inspired by these ideas, in this section, we develop a continuous-variable discrimination scheme, which we refer to as \textit{information recycling} (IR), for $N>2$ uniformly sampled phase-symmetric coherent states. The IR protocol combines optimal UD with a subsequent MED measurement on the failure states. In the following, we explicitly determine the structure of the failure states and characterize the performance of the IR discrimination. This strategy provides deterministic state discrimination together with a subset of error-free results obtained with the optimal success probability (\ref{eq:Ps}). These error-free results are still identifiable through the ancillary system [see Eq.~(\ref{eq:unitary})]. Finally, we use the mutual information between the random variables associated with state preparation and measurement to quantify the residual information encoded in the set of failure states.

\begin{figure*}[t]
\includegraphics[width=\textwidth]{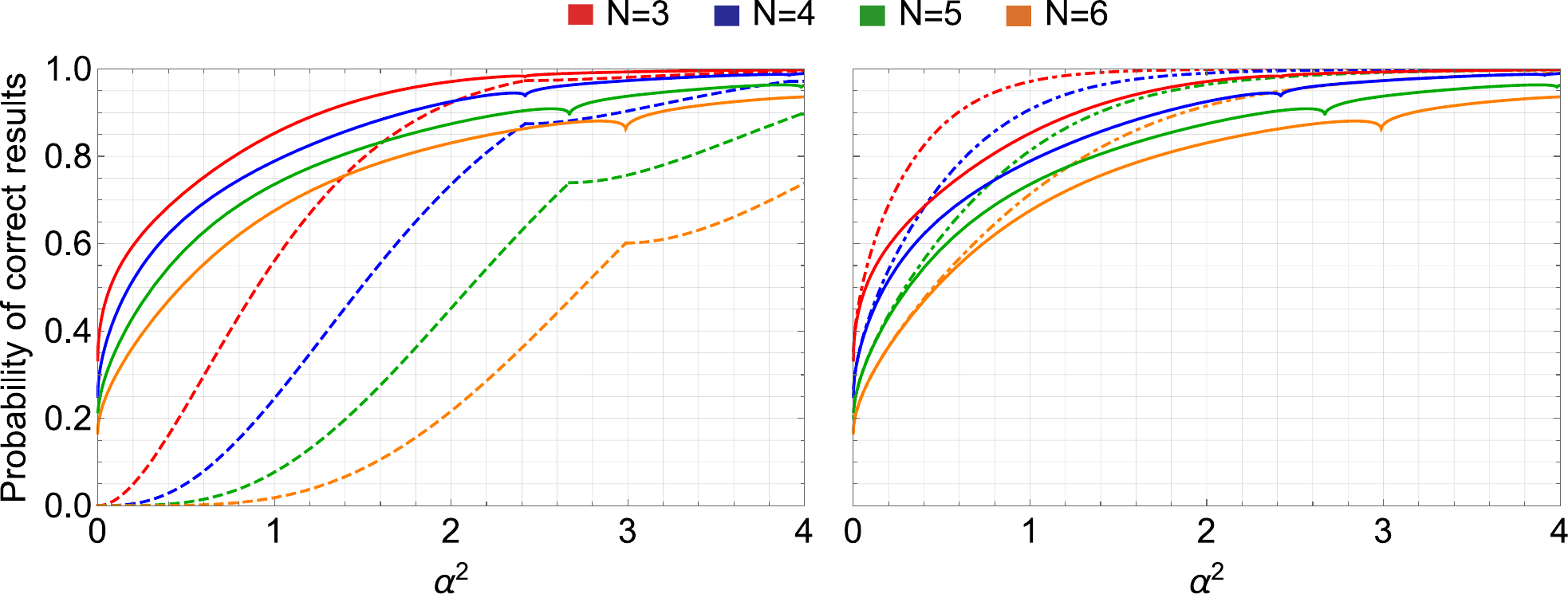}
\caption{Average probabilities of correct results for the IR strategy (solid lines in both panels) [Eq.~(\ref{eq:Pc})], optimal UD (dashed lines on the left panel) [Eq.~(\ref{eq:Ps})], and MED (dot-dashed lines on the right panel) [Eq.~(\ref{eq:Helstrom})] as functions of the mean photon number for sets with $N=3$ (red), $4$ (blue), $5$ (green) and $6$ (orange) states. IR outperforms optimal UD with respect to average probability of correct identifications, but it is, of course, outperformed by MED. The kinks in the plots correspond to points where the multiplicity of $c_\textrm{min}$ is greater than one (see text for details).} \label{fig:probabilities}
\end{figure*}

\subsection{POVM elements and detection probabilities} \label{subsec:recycling}

We consider and analyze the case where one performs a MED on the failure state set $\{|\beta_k\rangle\}$ of the state separation map~(\ref{eq:unitary}) in order to optimally extract the residual information about the signal states. The IR strategy is represented by a ($2N$)-outcome POVM with elements $\lbrace \hat{\Pi}^s_0,\ldots,\hat{\Pi}^s_{N-1} , \hat{\Pi}^f_0 ,\ldots,\hat{\Pi}^f_{N-1}\rbrace$, where $\sum_{j=0}^{N-1} (\hat{\Pi}^s_j + \hat{\Pi}^f_j ) = \hat{I}$ and $\hat{\Pi}^f_j  = \hat{A}_j^{f~\dagger} \hat{A}^f_j$ with the Kraus operators [see Eq.~(\ref{eq:ME_POVM})] being given by 
\begin{align}
    \hat{A}^f_j  &= \sqrt{\hat{\Pi}_j^{\textsc{med}}}\hat{A}^f  \label{eq:IRAf}.
\end{align}
The states in the failure set $\lbrace |\beta_k\rangle \rbrace_{k=0}^{N-1}$ [see Eq.~(\ref{eq:unitary})] can be easily calculated using the Kraus operator~(\ref{eq:SepAf}):
\begin{align}
    |\beta_k\rangle &= \frac{\hat{A}^f |\alpha_k\rangle}{\sqrt{1-P_s}} \label{eq:krausbtilde} \\
    &= \sum_{j=0}^{N-1} b_j \omega^{kj} \hat{U}_f |\phi_j\rangle, \label{eq:betaktilde}
\end{align}
where
\begin{align} \label{eq:bktilde}
    b_j = \sqrt{\frac{c_j^2-P_s/N}{1-P_s}}.
\end{align}
As expected, if $P_s\rightarrow 0$, then $b_j\rightarrow c_j$ and $|\beta_j\rangle\rightarrow |\alpha_j\rangle$. Note that, as evidenced by Eq.~(\ref{eq:betaktilde}), the failure states inherit the same mathematical structure of the signal states~(\ref{eq:coherent}) for a proper choice of $\hat{U}_f$. For this reason, we choose $\hat{U}_f=\hat{I}$ and, as a result, the failure set is also symmetric; however, they span a Hilbert space with fewer dimensions than the original and, therefore, are linearly dependent~\cite{Prosser16}. With this choice of $\hat{U}_f$, states~(\ref{eq:betaktilde}) can be discriminated with the minimum-error projectors defined in Eq.~(\ref{eq:ME_POVM}). In this case, it is still possible to obtain correct identifications of the signal state with probability:
\begin{align}   
    P^{\textsc{med},\beta}_c
    &=\frac{1}{N}\left(\sum_{j=0}^{N-1} b_{j}\right)^2, \label{eq:p_c_beta}
\end{align}
which corresponds to the Helstrom bound for equiprobable symmetric states [see Eq~(\ref{eq:Helstrom})], but now for the discrimination of the failure states. Consequently, the average probability of correct identifications in IR is given by
\begin{align}     \label{eq:Pc}
    P_c^\textsc{ir} = P_s + (1-P_s)P^{\textsc{med},\beta}_c.
\end{align}
Note that the probability in Eq.~(\ref{eq:p_c_beta}) is greater than a random guess, i.e. $P^{\textsc{med}, \beta}_c \geq 1/N$, except in situations where the multiplicity of $c_\textrm{min}$ is $\mu(c_\textrm{min}) = N-1$. This condition naturally excludes the binary signal ($N=2$) as a case of interest since, in this case, the failure space is one-dimensional ($|\beta_0\rangle = |\beta_1\rangle$) and there is no residual information that could be used for further discrimination.

According to Eq.~(\ref{eq:Pc}), the average probability of correct identifications for IR outperforms optimal UD by an amount equal to $(1-P_s)P_c^{\textsc{med},\beta}$. Naturally, this improvement will depend on the amplitude of the input coherent states. In Fig.~\ref{fig:probabilities}, we plot the average probability of correct results for IR [Eq.~(\ref{eq:Pc})] (solid lines on both panels), optimal UD [Eq.~(\ref{eq:Ps})] (dashed lines on the left panel), and MED [Eq.~(\ref{eq:Helstrom})] (dot-dashed lines on the right panel) as functions of the mean photon number. As expected, all probabilities tend to $1$ in the limit $\alpha \rightarrow \infty$, where the signal states are distinguishable. The performance gap between IR and optimal UD widens with increasing number of states for a fixed mean photon number. This reflects the fact that UD produces more inconclusive outcomes for larger sets, leaving more information to be recovered by the subsequent MED stage.  The difference between MED and IR is non-monotonic, vanishing in both the low-amplitude ($\alpha\rightarrow 0$) and high-amplitude ($\alpha\rightarrow\infty$) regimes. This behavior reflects the fundamental trade-off inherent to IR: By incorporating UD as the first stage, the protocol sacrifices some of MED's overall accuracy in exchange for a subset of unambiguous identifications. For instance, with three states at $\alpha^2=0.8$ in Fig.~\ref{fig:probabilities}, implementing IR instead of pure MED increases the average error probability by approximately $15\%$, while providing a $42\%$ success rate for error-free outcomes. Such a trade-off could be potentially relevant in CV-QKD applications, for instance, where distinguishing reliable from unreliable data could possibly be used to reduce post-processing costs.

To fully characterize this balance between accuracy and ambiguity, we now examine the individual confidences achieved by IR. The probabilities of obtaining the outcome $k'$ by measuring the system for success and failure events, represented by symbols $s$ and $f$ respectively, given that the signal state was $|\alpha_k\rangle$ are given by
\begin{subequations} \label{eq:ptilde}
\begin{align}
    p(k',s|\alpha_k) &= P_s \delta_{k'k} \label{eq:pk's|ak} \\
    p(k',f|\alpha_k) &= (1-P_s)\frac{1}{N} \sum_{m,\ell=0}^{N-1} \omega^{(k'-k)(\ell-m)} b_\ell b_m. \label{eq:pk'f|ak}
\end{align}
\end{subequations}
The prior probability of obtaining the outcome $k'$ is 
\begin{align}\label{eq:pk_prime}
    p(k')=\frac{1}{N} \sum_{k=0}^{N-1} \sum_{\ell=s,f} p(k',\ell|\alpha_k) = \frac{1}{N}, 
\end{align}
reflecting the fact that the input states are equiprobable. Using Bayes' rule, it is easy to show that the individual confidences [Eq.~(\ref{eq:confidence})] for success and failure events in the IR strategy are, respectively, given by
\begin{subequations}  \label{eq:IRconfs}
    \begin{align}
        C_k^s &\equiv p(\alpha_k|k,s)=1 \label{eq:Cs} \\
        C_k^f &\equiv p(\alpha_k|k,f)=P^{\textsc{med},\beta}_c. \label{eq:Cf}
    \end{align}
\end{subequations}
Although the average confidence is equal to Eq.~(\ref{eq:Pc}), an error-free set, which is identifiable by the ancillary system, is obtained with the optimal probability given by Eq.~(\ref{eq:Ps}).

\subsection{Failure states: Disturbance due to state separation}\label{subsec:failureset}

As shown in Eq.~(\ref{eq:bktilde}), the failure outputs of state separation [Eq.~(\ref{eq:unitary})] span an [$N-\mu(c_\textrm{min})$]-dimensional space, where $\mu(c_\textrm{min})$ is the multiplicity of $c_\textrm{min}$, so the number of dimensions lost in the transformation is ultimately a function of the mean photon number of the signal states, i.e. $\mu = \mu(\alpha^2)$. A consequence of the dimension reduction of the failure space is that the states in the failure set are less distinguishable than the input states, i.e. $|\langle \beta_j|\beta_k\rangle| > |\langle \alpha_j|\alpha_k\rangle|$ for $j\neq k$. This can be shown by direct calculation of the overlaps from Eq.~(\ref{eq:krausbtilde}), which yields
\begin{align}    \label{eq:overlapb}
   \langle \beta_j|\beta_k\rangle = \frac{\langle \alpha_j|\alpha_k\rangle}{1-P_s},
\end{align}
where we used the completeness relation $\hat{A}^{f~\dagger}\hat{A}^f=\hat{I}-\hat{A}^{s~\dagger}\hat{A}^s$. Despite being more similar to each other, each state in the failure set can still be associated with the signal state that has the same phase. Using the triangle inequality, we can show that the magnitude of the overlap,
\begin{align}  \label{eq:overlapab}
   |\langle \alpha_j|\beta_k\rangle| = \left|\sum_{\ell= 0}^{N-1} c_\ell b_\ell\omega^{\ell(k-j)}\right| ,
\end{align}
is maximized for $j=k$, i.e. $|\langle \alpha_j|\beta_k\rangle| \leq |\langle \alpha_k| \beta_k\rangle|$. Therefore, detecting the state $|\beta_k\rangle$ provides a reasonable inference that the original system state was $|\alpha_k\rangle$. We quantify the disturbance introduced by the state separation map through the infidelity $1-F$, where $F\equiv|\langle\alpha_k| \beta_k\rangle|^2$, between the signal state and the corresponding failure state.

Although an exact evaluation of Eqs.~(\ref{eq:coefficients}) and (\ref{eq:bktilde}) requires numerical methods, analytical insight can be gained in the regime where $c_\textrm{min}$ is much smaller than the remaining coefficients. For $c_j\neq c_\textrm{min}$ (otherwise, $b_j=0$), coefficients $\lbrace b_j\rbrace_{j=0}^{N-1}$ [Eq.~(\ref{eq:bktilde})] can be expanded as
\begin{align}
    b_j = \frac{c_j}{\sqrt{1-P_s}}\left[\left(1-\frac{c_\textrm{min}^2}{2c_j^2}\right) + \mathcal{O} \left(\frac{c_\textrm{min}^4}{c_j^4}\right)\right].
\end{align}
In the low mean photon number regime ($\alpha^2 \ll 1$), we can take the lowest-order approximation of the equation above, for $c_\textrm{min} \ll c_j~\forall~c_j\neq c_\textrm{min}$ (see Fig.~\ref{fig:coefficients}). In this regime, the multiplicity of the minimum coefficient $\mu(c_\textrm{min})=1$ as can be seen in Fig.~\ref{fig:coefficients}. Within this approximation, the fidelity $F$ becomes:
\begin{align}
    F &\approx \frac{1}{1-P_s}\left[\sum_{\ell=0}^{N-1}\left( c_\ell^2-\frac{c_\textrm{min}^2}{2}\right)(1-\delta_{c_\ell}^{c_\textrm{min}})\right]^2 \nonumber\\
    &= \frac{1}{1-P_s}\left[1-\frac{(N+1)}{2 N} P_s\right]^2,  \label{eq:overlapPs}
\end{align}
where we used $\sum_\ell c_\ell^2(1-\delta_{c_\ell}^{c_\textrm{min}}) = 1- c_\textrm{min}^2$ and the fact that the above sum has $N-1$ elements. Then it can be shown that the success probability is lower bounded by the infidelity between the input and its correspondent failure output $P_s \geq 1 - F$. The infidelity between the signal state and the corresponding failure state also imposes a lower bound on the minimum probability of error for discrimination between the failure output states, $P^{\textsc{med},\beta}_e=1-P^{\textsc{med},\beta}_c$ [see Eq.~(\ref{eq:p_c_beta})]. To see this, we take the square of Chebyshev's sum inequality,\footnote{Chebyshev's sum inequality, which consists of the bound
\begin{align}
    \left(\sum_{j=0}^{N-1} c_j \right) \left(\sum_{j=0}^{N-1} b_j \right) \leq N \left(\sum_{j=0}^{N-1} c_jb_j \right), \nonumber
\end{align}
holds as long as the coefficients have same ordering, i.e. $c_0\geq c_1 \geq \ldots \geq c_{N-1}$ and $b_0\geq b_1 \geq \ldots \geq b_{N-1}$. Note that, due to the relation (\ref{eq:bktilde}), this is the case since the sets $\lbrace b_j \rbrace_{j=0}^{N-1}$ and $\lbrace c_j \rbrace_{j=0}^{N-1}$ have the same ordering: $c_k\geq c_{\ell}$ immediately implicates that $b_k\geq b_\ell$ for any $k,\ell=0,\ldots,N-1$.} and using Eqs.~(\ref{eq:Helstrom}), (\ref{eq:p_c_beta}), and (\ref{eq:overlapab}), we straightforwardly obtain:
\begin{align}    \label{eq:bound}
    P^{\textsc{med},\beta}_e \geq 1- \frac{F}{P^\textsc{med}_c}.
\end{align}
Since $0\leq P^\textsc{med}_c \leq 1$, the error probability for the discrimination of the failure states is never smaller than the infidelity between the signal and failure states, i.e. $P^{\textsc{med},\beta}_e\geq 1 - F$.

\begin{figure}[t]
\includegraphics[width=\columnwidth]{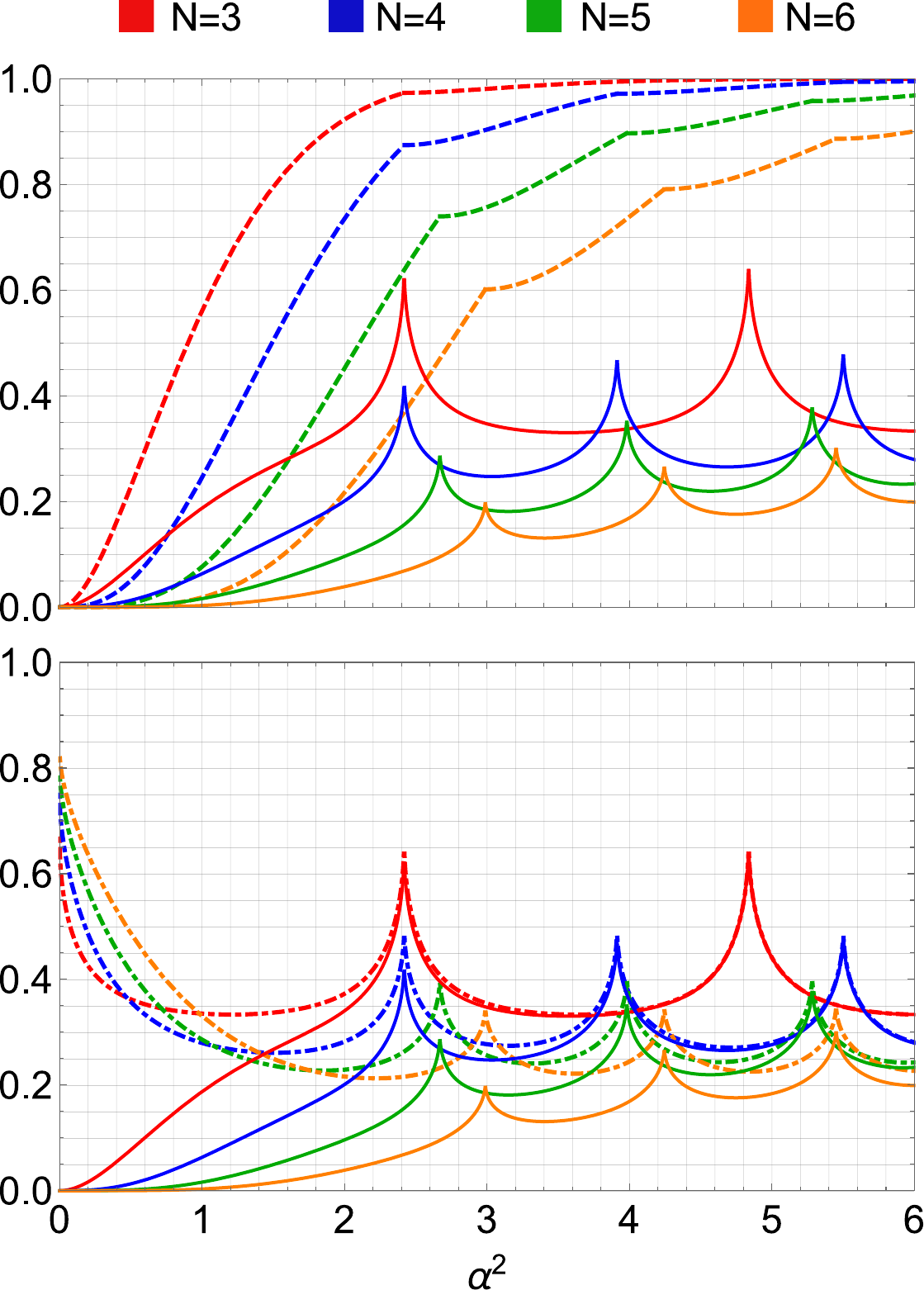}
\caption{(top) Success probability of optimal UD (dashed lines), Eq.~(\ref{eq:Ps}), and (bottom) error probability for the failure states (dash-dotted lines), $P^{\textsc{med},\beta}_e=1-P^{\textsc{med},\beta}_c$ [Eq.~(\ref{eq:p_c_beta})], as a function of mean photon number, $\alpha^2$, for sets of $N=3$ (red), $4$ (blue), $5$ (green) and $6$ (orange) phase-symmetric coherent states. For reference, we also plot in both graphs the infidelity between signal and failure states (solid lines) [Eq.~(\ref{eq:overlapab})], which quantify the disturbance of the state separation map on the input states, in case of failure. The kinks in all plots correspond to situations where the multiplicity of $c_\textrm{min}$ is greater than one (see text for details): In Appendix~\ref{app:analytical}, we show how to determine the corresponding values of $\alpha^2$ analytically. Note that the error probability for the failure states becomes equal to the infidelity as $\alpha^2$ increases.} \label{fig:infidelity}
\end{figure}

Figure~\ref{fig:infidelity} presents two complementary views of the IR strategy performance. The upper graph shows the optimal success probability~(\ref{eq:Ps}) as a function of the mean photon number $\alpha^2$, while the lower graph displays the minimum probability of error conditioned on failure outcomes, i.e. $P^{\textsc{med},\beta}_e = 1-P^{\textsc{med},\beta}_c$ [see Eq.~(\ref{eq:p_c_beta})], as a function of $\alpha^2$. In both graphs, we also plot the infidelity $1-|\langle\alpha_k|\beta_k\rangle|^2$, which quantifies the disturbance of the state separation map on the input states. The upper plot reveals a favorable operating regime where the separation map achieves non-negligible success rates at the cost of introducing low disturbance. For instance, from the upper plot, we see that for $1-|\langle \alpha_k|\beta_k\rangle|^2\approx 0.05$, $P_s\approx 0.15, 0.20, 0.25, 0.30$ for $N=3,4,5,6$, respectively. However, the complete trade-off becomes apparent only when considering the error probability in the lower graph. The error rate exhibits pronounced peaks near the kinks, which correspond to degeneracies in the minimum coefficient $c_\textrm{min}$. At these points, the failure space dimension drops below $N-1$, making discrimination of failure states more challenging. Despite this effect, by comparing both graphs, one can identify regions of efficient information recycling. As an example, we consider the $N=3$ case. For $0.6\lesssim \alpha^2 \lesssim 1.8$, the error probability exhibits a flat behavior, being approximately equal to $35\%$, while the success probability ranges approximately from $0.3$ to $0.9$. The bottom plot of Fig.~\ref{fig:infidelity} also shows that the bound~(\ref{eq:bound}) becomes tighter as the amplitude of the signal states increases (where $P^\textsc{med}_c \rightarrow1$); in this regime, the probability of error for the failure states becomes equal to the infidelity. 

As we mentioned in the previous paragraph, the kinks in all curves in Fig.~\ref{fig:infidelity} correspond to points where the minimum coefficient $c_\textrm{min}$ becomes degenerate. In Appendix~\ref{app:analytical}, we show how to analytically determine the values of the mean photon number for which the degeneracies occur by adapting the method introduced in~\cite{Bergou12} to continuous-variable states. As an example, we consider the three-state problem, for which we show that the kinks are periodic, and we analytically determine their period. In that case, all kinks correspond to situations where the failure space is one-dimensional, so there is no residual information on the failure set that could be used for discrimination by further measurements.

\subsection{Information gain}

In this section, we quantify the information-theoretic performance of IR. Specifically, we compute the classical mutual information provided by optimal UD and IR strategies and determine the information gain obtained with the recycling stage.

Let us start with the definition of classical mutual information in a generic prepare-and-measure communication scenario. Let $\mathcal{P}$ be the random variable corresponding to the label of the state that was prepared, so it takes values from the set $\lbrace 0,\ldots,N-1\rbrace$, and let $\mathcal{M}$ be the random variable associated with the label of the measurement outcomes, with $|\mathcal{M}|\geq |\mathcal{P}|$. The form and size of $\mathcal{M}$ depend on the measurement scheme chosen. For optimal UD, we have $|\mathcal{M}|=N+1$ since the labels of the measurement results belong to the set $\lbrace 0,\ldots,N-1,? \rbrace$, where $?$ represents the inconclusive result. On the other hand, the IR strategy requires $|\mathcal{M}|=2N$ outcomes, since their labels belong to the set $\lbrace (0,s),\ldots, (N-1,s), (0,f),\ldots,(N-1,f)\rbrace$. 

The classical mutual information between $\mathcal{P}$ and $\mathcal{M}$ is defined as
\begin{align}   \label{eq:mutual}
    I(\mathcal{P}:\mathcal{M}) = H(\mathcal{P}) - H(\mathcal{P}|\mathcal{M}),
\end{align}
where $H(\lbrace p_x \rbrace) = - p_x \log_2 p_x$ is the Shannon entropy and
\begin{align}  \label{eq:conditional}
    H(\mathcal{P}|\mathcal{M}) = \sum_{k'=0}^{N-1} \sum_{\ell\in\lbrace s,f\rbrace} p(k',\ell) H(\mathcal{P}|\mathcal{M}=k',\ell),
\end{align}
is the conditional entropy. Here, $p(k',\ell)$ corresponds to the joint probability distribution for system outcome $k'$ and ancilla outcome $\ell$. From this point forward, we will use the less cumbersome notation $H(\mathcal{P}|k',\ell) =H(\mathcal{P}|\mathcal{M}=k',\ell)$. Due to the symmetry of the failure set, we can rewrite Eq.~(\ref{eq:conditional}) as
\begin{align}
   H(\mathcal{P}|\mathcal{M}) &= N\sum_{\ell\in\lbrace s,f\rbrace} p(0,\ell) H(\mathcal{P}|0,\ell) \nonumber\\
   &=\sum_{\ell\in\lbrace s,f\rbrace} p(\ell) H(\mathcal{P}|0,\ell) \nonumber \\
   &= (1-P_s) H(\mathcal{P}|0,f), \label{eq:conditional2}
\end{align}
where we used that $p(k',\ell)=p(\ell)/N$ [see Eq.~(\ref{eq:pk_prime})] and $H(\mathcal{P}|k',s)=0$ for all $k'$, since in the case of success $p(\alpha_k|k',s) = \delta_{k,k'}$ [see Eqs.~(\ref{eq:Bayes}) and (\ref{eq:pk's|ak})]. Substituting Eq.~(\ref{eq:conditional2}) into Eq.~(\ref{eq:mutual}), and using that $H(\mathcal{P})= \log_2 N$, the mutual information for the IR strategy yields
\begin{align}    \label{eq:mutualIR}
    I^\textsc{ir} (\mathcal{P}:\mathcal{M}) = \log_2 N - (1-P_s) H(\mathcal{P}|0,f),
\end{align}
where $H(\mathcal{P}|0,f)$ is the Shannon entropy of the random variable associated with the detection of the failure states, with probabilities given by [see Eqs.~(\ref{eq:Bayes}) and (\ref{eq:pk'f|ak})]
\begin{align}
    p(\alpha_k|0,f) = \frac{1}{N} \sum_{m,\ell=0}^{N-1} \omega^{-k(\ell-m)} b_\ell b_m. 
\end{align}

As for optimal UD, the mutual information is simply $I^\textsc{ud}(\mathcal{P}:\mathcal{M})=P_s\log_2 N$, so that the information gain obtained through the recycling stage is
\begin{align}\label{eq:info_gain}
  I^\textsc{ir} (\mathcal{P}:\mathcal{M}) - I^\textsc{ud}(\mathcal{P}:\mathcal{M}) = (1-P_s)\left[\log_2 N - H(\mathcal{P}|0,f)\right].
\end{align}

\begin{figure}[t]
\includegraphics[width=\columnwidth]{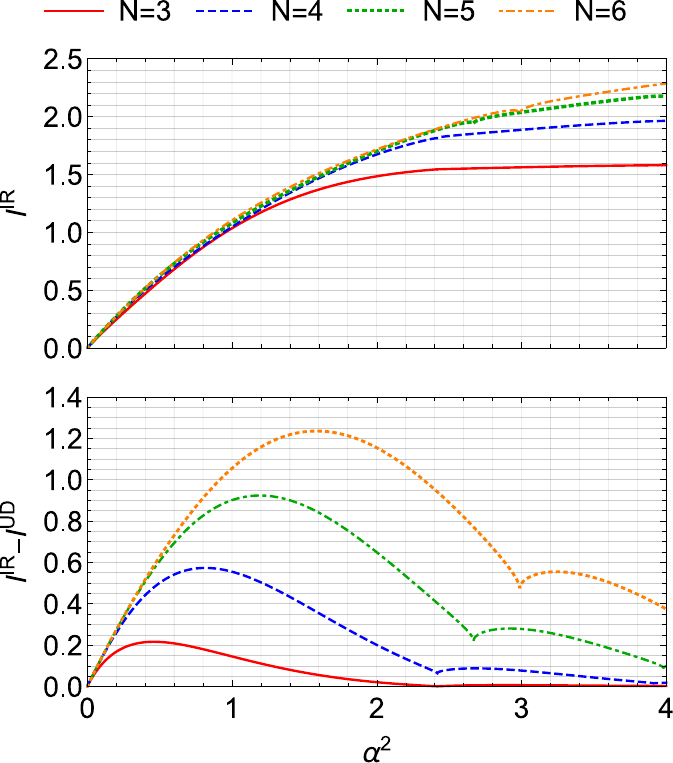}
\caption{(top) Mutual information of the IR strategy [Eq.~(\ref{eq:mutualIR})] and (bottom) information gained in the recycling stage [Eq.~(\ref{eq:info_gain})], both as a function of the mean photon number $\alpha^2$ for sets of $N=3$ (solid), $4$ (dashed), $5$ (dot-dashed) and $6$ (dotted) phase-symmetric coherent states. Notably, for $\alpha^2 \lesssim 1$, $I^\textsc{ir}(\mathcal{P}:\mathcal{M})$ shows minimal $N$-dependence (top), yet recycling yields substantial improvements over UD (bottom). Not surprisingly, the increase of the information gain with $N$ is due to the fact that optimal UD discards progressively more information for larger $N$. As $\alpha^2$ increases and the signal states become more distinguishable, the information gained by recycling of the failure states decreases. The kinks in all plots correspond to situations where the multiplicity of $c_\textrm{min}$ is greater than one (see text for details).} \label{fig:mutual}
\end{figure}

Note that from Eq.~(\ref{eq:mutualIR}), it follows that $H(\mathcal{P}|0,f) \leq\log_2 N$. Thus, from Eq.~(\ref{eq:info_gain}), we have that $I^\textsc{ir}(\mathcal{P}:\mathcal{M})\geq I^\textsc{ud}(\mathcal{P}:\mathcal{M})$. The equality $I^\textsc{ir}(\mathcal{P}:\mathcal{M}) = I^\textsc{ud}(\mathcal{P}:\mathcal{M})$ holds only when there is no residual information in the failure states, which occurs for three trivial scenarios: $(i)$ $\alpha=0$ (vacuum), $(ii)$ $\alpha \rightarrow \infty$ (orthogonal signal states) and $(iii)$ a one-dimensional failure space [$H(\mathcal{P}|0,f)=\log_2 N$]. 

Figure~\ref{fig:mutual} shows the IR protocol's information-theoretic performance; we plot the mutual information provided by the IR strategy~(\ref{eq:mutualIR}) (upper graph) and the information gained from recycling the failure states~(\ref{eq:info_gain}) (lower graph) both as functions of the mean photon number, $\alpha^2$, for $N=3$ (solid), $4$ (dashed), $5$ (dot-dashed) and $6$ (dotted). As expected, the mutual information approaches $\log_2 N$ as $\alpha\rightarrow\infty$, where states become distinguishable. For $\alpha^2\lesssim1$, $I^{\textsc{ir}}$ shows little variation for different $N$ (upper graphs). Despite this behavior, the recycling stage delivers substantial improvements over optimal UD alone, as evidenced by the significant information gain shown in the lower graph. In this regime, optimal UD discards more information as $N$ increases--information that is successfully recovered by IR. As expected, the information exhibits nonmonotonic behavior with $\alpha^2$, since states become more distinguishable with increasing $\alpha^2$, reducing the benefit from the recycling stage. IR achieves its peak performance at $\alpha^2_{\textrm{opt}} \approx 0.4, 0.8, 1.2, 1.6$ for $N=3,4,5,6$, respectively. The kinks in the lower graph correspond to drops in information gain caused by $c_{\textrm{min}}$ degeneracies that reduce the failure space dimension, degrading the discriminability of failure states. As shown in Fig.~\ref{fig:infidelity}, these points coincide with peaks in the infidelity between input and failure states.

\section{Conclusion}\label{sec:conclusion}

In this paper, we have revisited the problem of optimal UD for a set of equiprobable $N$ phase-symmetric coherent states and demonstrated that, for $N>2$, substantial residual information remains in the failure states of the state separation map. While these failure states cannot yield additional unambiguous identifications, they retain useful information for further discrimination that can be optimally extracted via MED. By recycling failure states, the discrimination protocol becomes deterministic while preserving a subset of error-free identifications with optimal success probability. This IR strategy opens up new possibilities for adaptive and sequential discrimination protocols in continuous-variable settings, and suggests that measurement failures--often viewed as inconclusive results--can be leveraged as a resource for extracting additional information. 

\begin{figure*}[tbp]
\centering
\includegraphics[width=\textwidth]{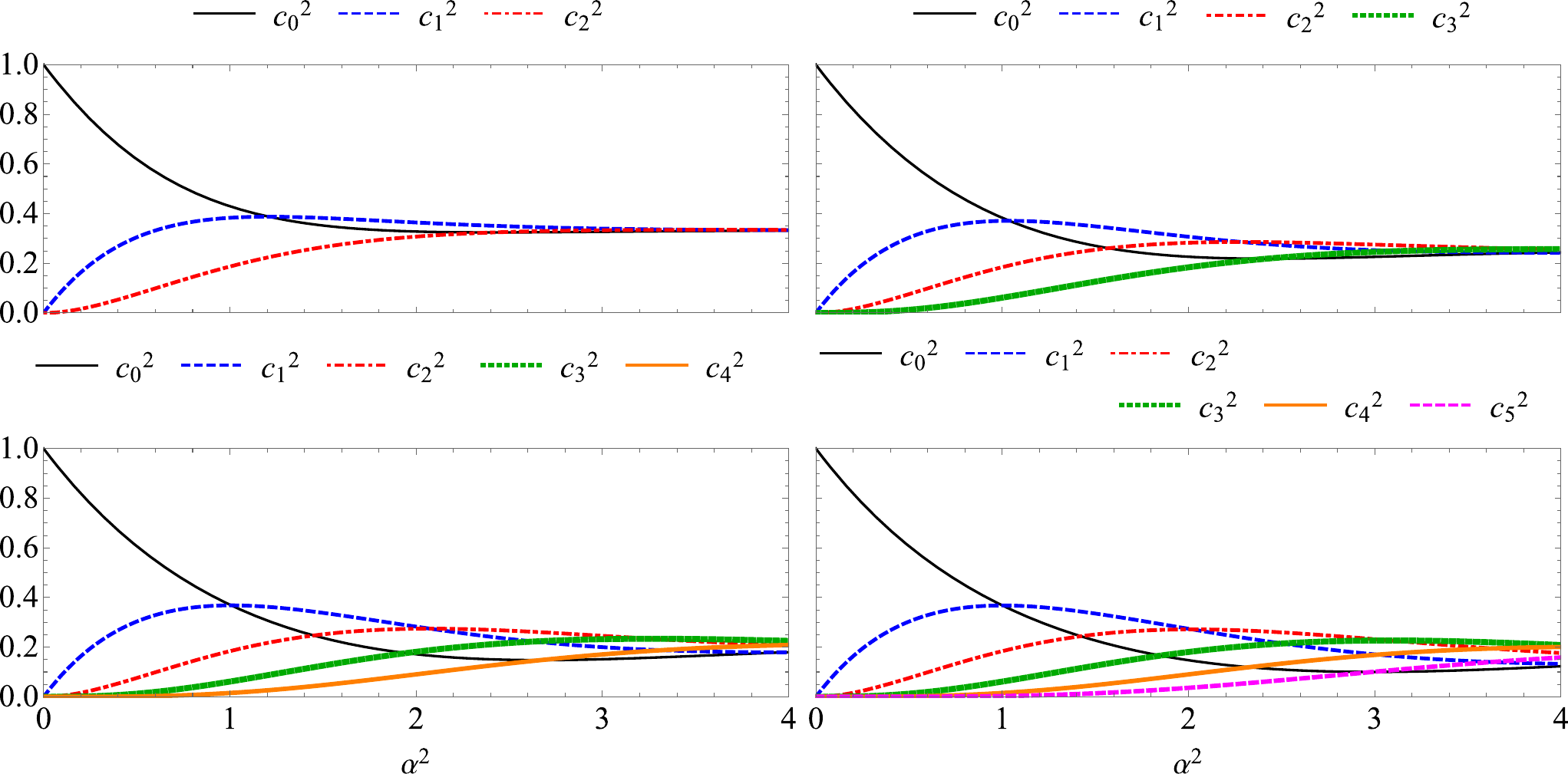}
\caption{Squared coefficients $\lbrace c_j^2 \rbrace_{j=0}^{N-1}$ [Eq.~(\ref{eq:coefficients})] as functions of the mean photon number for phase-symmetric sets with $N=3$ (left plot on first row), $4$ (right plot on first row), $5$ (left plot on second row) and $6$ (right plot on second row) coherent states.}
\label{fig:coefficients}
\end{figure*}

We characterized protocol performance through complementary probabilistic and information-theoretic analyses. We considered the infidelity between the input and failure states as a measure of the disturbance introduced by the state separation map. We showed that it provides a lower bound on recycling-stage errors, remaining relatively small in favorable operating regimes where IR achieves significant unambiguous identification rates. In addition, we calculated the classical mutual information for both IR and optimal UD, revealing substantial information gains from recycling: IR systematically recovers the residual information left in UD failure states, which increases for larger state alphabets.

These features make IR a particularly promising non-conventional receiver strategy for phase-shift keying CV-QKD protocols, where post-processing costs can be a limiting factor~\cite{Pirandola20}. By flagging reliable (error-free) outcomes via an ancillary degree of freedom, IR enables adaptive postprocessing strategies: Unambiguous results bypass correction entirely, while recycled data undergo processing for the failure ensemble. This approach could potentially reduce computational overhead in practical implementations, which motivates further research.

\section*{Acknowledgments}

This work has been partially funded by the project ``Receptores não-convencionais em CV-QKD'' supported by the EMBRAPII CIMATEC Competence Center in Quantum Technologies - Quantum Industrial Innovation (QuIIN), with financial resources from the PPI IoT/Manufatura 4.0 of the MCTI grant number 053/2023, signed with EMBRAPII. The research of J. A. B. was supported by NSF Grant No. 2504622.

\appendix

\section{Orthonormal basis} \label{app:orthonormal}

As discussed in Sec.~\ref{subsec:symmetric}, we can expand the signal states (\ref{eq:coherent}) on the orthonormal basis $\lbrace |\phi_j\rangle \rbrace_{j=0}^{N-1}$ [see Eq.~(\ref{eq:alphacj})], which allows us to determine the POVM operators corresponding to the optimal discrimination strategies addressed in this paper. In this section, we derive the explicit expression for the states in this basis. 

Consider the expansion of the states $\lbrace|\phi_j\rangle\rbrace$ on the Fock basis $|\phi_j\rangle = \sum_{n=0}^\infty \phi^j_n|n\rangle$. Comparing Eqs.~(\ref{eq:coherent}) and (\ref{eq:alphacj}), we find that coefficients $\phi_n^j$ must satisfy the relation:
\begin{align}  \label{eq:A1}
    \sum_{j=0}^{N-1} c_j \omega^{k(j-n)} \phi^j_n  = e^{-\alpha^2/2}  \frac{\alpha^n}{\sqrt{n!}},
\end{align} 
where $\omega=e^{2\pi i/N}$ is the $N$th root of unity, $n\in\mathbb{N}$, and the coefficients $\lbrace c_j\rbrace$ are given by Eq.~(\ref{eq:coefficients}). Since the left-hand side of Eq.~(\ref{eq:A1}) must have a null phase, the solution requires $n-j=pN$ with $p\in\mathbb{N}$, which leads to
\begin{align}  \label{eq:A2}
    \phi^j_n = \delta^n_{j+pN}~ \frac{e^{-\alpha^2/2}\alpha^n}{c_j\sqrt{n!}}.
\end{align}
As a result, the states $\lbrace |\phi_j\rangle\rbrace$ are given by
\begin{align}
    |\phi_j\rangle = \frac{e^{-\alpha^2/2}}{c_j} \sum_{p=0}^{\infty} \frac{\alpha^{j+pN}}{\sqrt{(j+pN)!}} |j+pN\rangle.
\end{align}
In Fig.~\ref{fig:coefficients}, we plot the squared coefficients $\lbrace c_j^2 \rbrace_{j=0}^{N-1}$ [see Eq.~(\ref{eq:coefficients})] as functions of the mean photon number for sets with $N=3,4,5,6$ phase-symmetric coherent states. Note that the approximation $c^2_\textrm{min}\ll c^2_j~\forall~c_j\neq c_\textrm{min}$ is valid in the low mean photon number regime ($\alpha^2\ll1$), and becomes more accurate as we increase the number of states in the set. In that regime, $c_\textrm{min}$ is non-degenerate. This approximation is used in Sec.~\ref{subsec:failureset}.

\begin{figure*}[t]
\includegraphics[width=\textwidth]{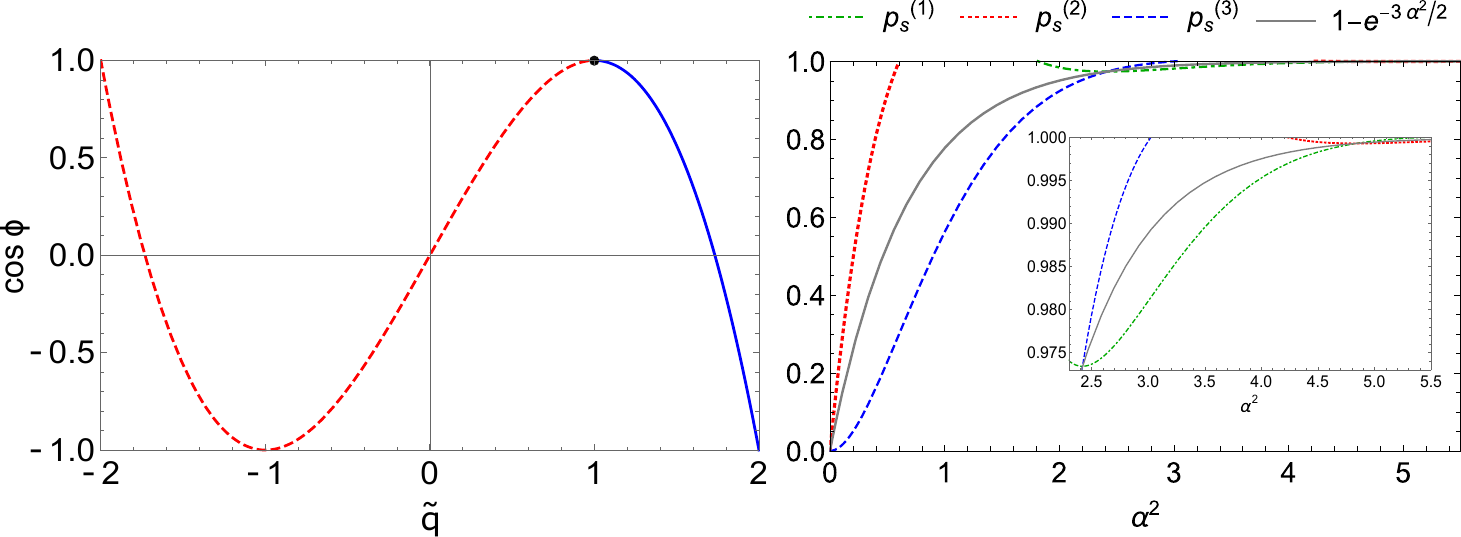}
\caption{Analytical solution to the optimal success probability of UD for a set with three phase-symmetric coherent states. Left: Plot of $\cos\phi$ [see Eq.~(\ref{eq:3rd})], where $\phi=3\sqrt{3} \alpha^2/2$ is the Berry phase. Since the failure probability must be larger than the overlap (see text for details), the branch of the function for $\tilde{q}<1$ (dashed red line) is non-physical. The black circle corresponds to the point where the failure space is one-dimensional (see text for details). Right: Solutions to the optimal success probability~(\ref{eq:solutions}) as functions of the mean photon number. The insert shows the points where the physical solution changes: From $p_s^{(3)}$ to $p_s^{(1)}$ at $\alpha^2= \frac{2\pi}{3\sqrt{3}}$, and from $p_s^{(1)}$ to $p_s^{(2)}$ at $\alpha^2= \frac{4\pi}{3\sqrt{3}}$.}
\label{fig:solution}
\end{figure*}

\section{Explicit analytical solution for the optimal success probability of unambiguous discrimination}\label{app:analytical}

In this appendix, we obtain analytical expressions for the optimal success probability of unambiguous discrimination of phase-symmetric coherent states using the method proposed in~\cite{Bergou12}, which we briefly describe below. This method additionally provides the exact values of the mean photon number for which the minimum coefficient degenerates, thus, the situations where the failure space has a lower dimension. As an example, we consider the three-state problem; we show that the failure space becomes one-dimensional at periodic values of mean photon number and determine the corresponding period.

As discussed in Sec.~\ref{subsec:recycling}, the failure states of the transformation (\ref{eq:unitary}) must be linearly dependent. This condition implies that $\text{det}(F)=0$, where $F_{jk}= \langle\beta_j| \beta_k\rangle$ corresponds to the Gram matrix of the failure states set. Using Eq.~(\ref{eq:krausbtilde}), such requirement can be written as:
\begin{align}   \label{eq:determinant}
    \text{det}(F)=\text{det}\left[\frac{G - p_sI}{1-p_s}\right]=0,
\end{align}
where $p_s$ is the success probability, $G_{jk}=\langle\alpha_j |\alpha_k\rangle$ is the Gram matrix of the signal states [see Eq.~(\ref{eq:innerprod})], and $I$ is the identity matrix of size $N$. For a set with $N$ states, the above equation yields an N$^{\textrm{th}}$ order polynomial equation that must be solved for $p_s$.

Without loss of generality, here we will discuss the example of three coherent states. Note that this set has an extra symmetry, since all overlaps are equal $|G_{01}|=|G_{12}|= |G_{20}|=\exp(-3\alpha^2/2)$ [see Eq.~(\ref{eq:innerprod})]. The authors of~\cite{Bergou12} showed that Eq.~(\ref{eq:determinant}) can be simplified to
\begin{align}  \label{eq:3rd}
    \tilde{q}^3 - 3\tilde{q} + 2\cos\phi=0,
\end{align}
where $\phi$ is the Berry phase and $\tilde{q}=(1-p_s)\exp(3\alpha^2/2)$ is the scaled failure probability. The Berry phase $\phi=\phi_0+\phi_1+ \phi_2$ is the phase deficiency corresponding to a closed path ($0\rightarrow 1 \rightarrow 2 \rightarrow 0$), where $\phi_j$ is the phase of complex overlap ($G_{01}= |G_{01}|e^{i\phi_2}$ and the two cyclic permutations). Using the inner product~(\ref{eq:innerprod}), we find $\phi=3\sqrt{3} \alpha^2/2$. The dependence of the Berry phase on the mean photon number highlights the nonlinearity of this problem for continuous-variable states (in the original paper, which addressed discrete-variable states, the authors were able to fix the Berry phase as an arbitrary constant). 

In Eq.~(\ref{eq:3rd}), the cosine function determines whether the equation has one or two nonnegative roots, as shown in the plot on the left-hand side of Fig.~\ref{fig:solution}. According to~\cite{Dieks88}, the failure probability of UD cannot be larger than the overlap, so the physical solution must be in the domain $\tilde{q}\geq1$ (represented in the plot by the solid blue line). As a result, the physical part of the solution can be represented by a point that goes downward and upward in the blue line as the mean photon number increases. The black circle in the figure corresponds to the point where solutions overlap, and the failure space is one-dimensional.

The solutions of Eq.~(\ref{eq:3rd}) yield
\begin{align}  \label{eq:solutions}
    \left\{\begin{array}{l} p_s^{(1)}= 1+ 2e^{-3\alpha^2/2}\cos\left(\frac{\sqrt{3}}{2}\alpha^2\right)\\[3mm]
    p_s^{(2)}=1- e^{-3\alpha^2/2}\left[\cos\left(\frac{\sqrt{3}}{2}\alpha^2\right) - \sqrt{3} \sin\left(\frac{\sqrt{3}}{2}\alpha^2\right)\right]\\[3mm]
    p_s^{(3)}=1- e^{-3\alpha^2/2}\left[\cos\left(\frac{\sqrt{3}}{2}\alpha^2\right) + \sqrt{3} \sin\left(\frac{\sqrt{3}}{2}\alpha^2\right)\right]
    \end{array}\right. .
\end{align}
These functions are plotted on the right-hand side of Fig.~\ref{fig:solution} in terms of the mean number of photons. Note that $p_s^{(1)}=p_s^{(2)}= p_s^{(3)}=0$ for the vacuum state ($\alpha=0$) and $p_s^{(1)}= p_s^{(2)}=p_s^{(3)}=1$ for distinguishable input states ($\alpha\rightarrow\infty$). For arbitrary values of $\alpha$, we must select the physical branch of the solutions in (\ref{eq:solutions}) by ensuring that it is continuous and smaller than $1-\exp(-3\alpha^2/2)$ (for more details, see Ref.~\cite{Dieks88}). In Fig.~\ref{fig:solution}, we show that this condition immediately selects one of the branches based on the mean number of photons (in the figure, the quantity $1-\exp(-3\alpha^2/2)$ is represented by the solid gray line). 

Note that the periodicity of the Berry phase determines the exact values of the mean photon number for which the physical solution changes among Eqs.~(\ref{eq:solutions}), which also corresponds to the situations where the failure space is one-dimensional. The periodicity is determined by the condition $\phi=2\pi m$ where $m$ is an integer, leading to a period of $\alpha^2=\frac{2\pi}{3\sqrt{3}}\approx 2.418$. As can be seen in the insert of the plot on the left-hand side of Fig.~\ref{fig:solution}, at this point, the physical solution changes from $p_s^{(3)}$ (dashed blue line) to $p_s^{(1)}$ (dot-dashed green line); and at $\alpha^2=\frac{4\pi}{3\sqrt{3}}\approx4.837$, the physical solution changes from $p_s^{(1)}$ to $p_s^{(2)}$ (dotted red line). The periodicity continues for $\phi>4\pi$; however, since all solutions are so close to $1$ beyond that point, they are practically indistinguishable.

Beyond providing an analytical expression for the optimal success probability~(\ref{eq:Ps}), this approach offers more physical insight about the structure of the failure states set, since it allows prediction of the mean photon number values where the failure space has a lower dimension. This method can be readily extended for larger sets of phase-symmetric coherent states.

\bibliographystyle{apsrev4-2}
\bibliography{references}

\end{document}